\documentclass[twocolumn,showpacs,preprintnumbers,amsmath,amssymb,pre]{revtex4}
\usepackage{graphicx}
\usepackage{dcolumn}% Align table columns on decimal point

\usepackage[centerlast,footnotesize]{caption}

\usepackage[latin1]{inputenc}
\usepackage[T1]{fontenc}
\usepackage{ae,aecompl}

\begin{document}
\captionsetup{labelsep= none,
  justification=centerlast,width=.9\columnwidth,aboveskip=5pt}

\newcommand{\rar}{$\rightarrow$}
\newcommand{\lrar}{$\leftrightarrow$}

\newcommand{\beq}{\begin{equation}}
\newcommand{\eeq}{\end{equation}}
\newcommand{\bea}{\begin{eqnarray}}
\newcommand{\eea}{\end{eqnarray}}
\newcommand{\Req}[1]{Eq. (\ref{E#1})}
\newcommand{\req}[1]{(\ref{E#1})}
\newcommand{\degree}{$^{\rm\circ} $}
\newcommand{\pcite}{\protect\cite}
\newcommand{\pref}{\protect\ref}
\newcommand{\Rfg}[1]{Fig. \ref{F#1}}
\newcommand{\rfg}[1]{\ref{F#1}}
\newcommand{\Rtb}[1]{Table \ref{T#1}}
\newcommand{\rtb}[1]{\ref{T#1}}

\title{\bf Local elasticity of strained DNA studied by all-atom
simulations}

\author{Alexey K. Mazur}
\email{alexey@ibpc.fr}
\affiliation{UPR9080 CNRS, Univ Paris Diderot, Sorbonne Paris Cité,\\
Institut de Biologie Physico-Chimique,\\
13, rue Pierre et Marie Curie, Paris,75005, France.}

\begin{abstract}
Genomic DNA is constantly subjected to various mechanical stresses
arising from its biological functions and cell packaging. If the local
mechanical properties of DNA change under torsional and tensional
stress, the activity of DNA-modifying proteins and transcription
factors can be affected and regulated allosterically. To check this
possibility, appropriate steady forces and torques were applied in the
course of all-atom molecular dynamics simulations of DNA with AT- and
GC-alternating sequences. It is found that the stretching rigidity
grows with tension as well as twisting. The torsional rigidity is not
affected by stretching, but it varies with twisting very strongly, and
differently for the two sequences. Surprisingly, for AT-alternating
DNA it passes through a minimum with the average twist close to the
experimental value in solution. For this fragment, but not for the
GC-alternating sequence, the bending rigidity noticeably changes with
both twisting and stretching. The results have important biological
implications and shed light upon earlier experimental observations.
\end{abstract}%.................................................


\maketitle

\section*{Introduction}

Internal mechanical stress is ubiquitous in the biologically active
state of double helical DNA. In eucaryotic cells, DNA is densely
packed in chromosomes and forced to bend, twist and stretch by
numerous protein factors involved in genome regulation
\cite{Bloomfield:00,Cozzarelli:06}. In procaryotes, DNA is subjected
to a constitutive unwinding torque maintained by special enzymes,
which leads to supercoiling, as in a long rope with bending and
twisting elasticity \cite{Wang:96,Wang:02a}. The supercoiling and,
more generally, stress-induced DNA forms are key factors in a
variety of cellular processes \cite{Vologodskii:94b}. For instance,
the degree of supercoiling in bacteria changes systematically during
the cell cycle and in response to environmental conditions, which is
accompanied by activation or suppression of certain genes
\cite{Travers:05a}. The promoter sensitivity to supercoiling stems
from the recognition of short promoter elements by RNA polymerase
\cite{Borowiec:87}. Detailed studies indicate that it probably does
not require DNA melting nor transitions to alternative forms
\cite{Travers:05a}. In {\em E. coli}, relaxation of the superhelical
stress simultaneously alters the activity of 306 genes (7\% of the
genome), with 106 genes activated and others deactivated
\cite{Peter:04}. The genes concerned are functionally diverse and
widely dispersed throughout the chromosome, and the effect is
dose-dependent.

The physical mechanisms of such effects are understood only partially.
Long DNA is well described by the coarse-grained worm-like chain (WLC)
model \cite{Landau:76,Cantor:80b} supplemented with harmonic twisting
and stretching elasticity
\cite{Bustamante:94,Vologodskii:94a,Vologodskii:97,Wang:97b,Moroz:97,Bouchiat:99}.
This model nicely explains the stress-modulated probability of
looping, wrapping around proteins, and juxtaposition of distant
protein binding sites \cite{Vologodskii:92}. However, it cannot
account for the promoter sensitivity to supercoiling, for instance,
because in this and many other cases the gene regulation has a strong
local character and is dominated by sequence effects. A long-discussed
hypothesis is that the stress may act as an allosteric factor in
protein-DNA recognition \cite{Wells:77,Bauer:78}. The supercoiling
arguably changes the local properties of DNA, as there are small
proteins with single short binding sites that can distinguish stressed
and relaxed DNA forms \cite{Balandina:02}; however, it is never clear
what exactly is recognized. The supercoiling torque is distributed
between twisting and writhing so that the untwisting of the double
helix is estimated as 1-2\% \cite{Boles:90}, which is below the
thermal noise and too small for reliable recognition. Alternatively,
the action of the torsional stress may be conveyed through a property
other than the structure of the double helix. For instance, the
untwisting may change the elastic parameters of DNA
\cite{Song:90b,Selvin:92,Naimushin:94}. The supercoiled DNA is
governed by the interplay between the local bending and twisting
fluctuations. If the bending flexibility or the torsional stiffness
vary, parameters of thermal fluctuations of short DNA stretches
involved in recognition could be noticeably affected even at low
levels of stress.

The foregoing hypothesis implies that even with small deformations
the DNA elasticity is not exactly harmonic. This possibility was
earlier considered in relation to specific experiments and also to
explain the discrepancies in twisting rigidity of DNA evaluated by
different methods
\cite{Song:90b,Selvin:92,Naimushin:94,Fujimoto:06}. Notably, it was
suggested that the stretching forces applied in single molecule
measurements and the bending involved in DNA cyclization can
increase the apparent twisting rigidity of DNA \cite{Fujimoto:06}.
The DNA double helix tends to overwind with small stretching
\cite{Gore:06b,Lionnet:06}, but it is not clear if bending and/or
stretching affect the twisting elasticity. The mechanical coupling
between deformations of different types may be very important for
regulation. However, the most interesting for biology is not the
overall elasticity, but the behavior of short specific sequences
within polymer DNA. To the present, all experimental studies have
probed only the average properties of long DNA, with a few reports
on sequence effects \cite{Fujimoto:90,Geggier:10} and the influence
of supercoiling stress \cite{Song:90b,Selvin:92,Naimushin:94}. For
the free relaxed double helix a good convergence of the results of
different experiments is obtained for the bending rigidity
\cite{Hagerman:88,Geggier:10}.  The torsional rigidity has been
measured by multiple different techniques, but the results remain
controversial \cite{Fujimoto:06}.  Also, a few estimates of the
stretching stiffness have been obtained from nanomechanics
experiments with single DNA molecules
\cite{Smith:96,Wang:97b,Wenner:02}.

Although the local sequence-dependent DNA elasticity and possible
stress effects are difficult to reveal experimentally, they can be
probed by computer simulations. All-atom molecular dynamics (MD)
simulations is a powerful instrument particularly suitable for this
purpose.  Continuous improvement of forcefields
\cite{Cornell:95,MacKerell:95,Perez:07a} and simulation techniques
\cite{Darden:93,Essmann:95} have now made possible free MD simulations
that reproduce conformational ensembles of DNA in good agreement with
experimental data \cite{Cheatham:00,Perez:07b}. Calculated statistics
of fluctuations in short DNA qualitatively agree with the WLC theory
\cite{Mzbj:06,Mzprl:07}, and the values of the elastic parameters can
be measured with good accuracy \cite{Mzjpc:08,Mzjpc:09}. DNA
deformation is a classical subject of molecular mechanics
\cite{Zhurkin:79}. In several earlier investigations, all-atom MD
simulations were used for studying deformed DNA states obtained by
external stretching \cite{MacKerell:99,Harris:05,Luan:08b}, twisting
\cite{Kannan:06,Wereszczynski:06,Randall:09}, or bending
\cite{Curuksu:08}. The required deformations were produced by either
potential restraints or periodical boundary constraints. A promising
alternative method \cite{Mzjctc:09} applies steady forces and torques
to short stretches of DNA. In contrast to the earlier approaches, this
method makes it possible to evaluate elastic parameters under
different types and magnitudes of external stress corresponding to
physiological conditions. This method captures linear elastic
responses as well as the twist-stretch coupling effect under small
torques corresponding to a physiological degree of supercoiling
\cite{Mzjctc:09}. With such approaches it has been found that,
depending upon the base pair sequence, small twisting torques
corresponding to physiological superhelical density can significantly
change the torsional stiffness of the DNA double helix
\cite{Mzprl:10}.

In this article we present the results of the first systematic study
of the influence of external mechanical stress upon the local
stretching, twisting, and bending elasticity of the double helical
DNA. The numerical algorithms described and tested in the recent
reports \cite{Mzjctc:09,Mzprl:10} could be drastically accelerated
through parallelization, which made such computations more
affordable.  Two double helical fragments were considered, with AT-
and GC-alternating sequences, respectively. We found that the
apparent stretching rigidity of DNA strongly depends upon the method
used for measuring the molecule length. When it is obtained by
summing base-pair steps as in earlier studies
\cite{Lankas:00,Mzbj:06,Mzjpc:09} the sign of the twist-stretch
coupling effect appears opposite to that measured experimentally. In
contrast, much better agreement with experimental data is obtained
when the length is measured directly via the end-to-end distance of
one helical turn. We argue that only the latter value corresponds to
the experimental observable. The change in the stretching rigidity
of DNA with external stress is qualitatively similar for the two
sequences.  It grows with stretching as well as with increased
twisting. The torsional rigidity is essentially unaffected by
stretching, but it varies with twisting very strongly, and
differently for the two sequences.  Surprisingly, for the
AT-alternating sequence, it passes through a minimum with the
average twist close to the experimental value in solution. For this
fragment, but not for the GC-alternating sequence, the bending
rigidity noticeably changes with both twisting and stretching. The
results shed light upon the earlier experimental observations
\cite{Song:90b,Selvin:92,Naimushin:94} and have important
implications for the possible mechanisms of allosteric gene
regulation \cite{Borowiec:87,Peter:04,Chow:91}.

\section*{Methods}

\section*{Simulation protocols}

Tetradecamer DNA fragments were modeled with AT-alternating and
GC-alternating sequences. A dodecamer fragment is necessary for a full
helical turn of a random-sequence B-DNA. The length of 14 base pairs
(bp) is minimal for modelling of a helical turn within a longer DNA.
This choice of the fragment length and sequences is consistent with
and dictated by the results of the earlier studies
\cite{Mzjpc:08,Mzjpc:09}. Steady stress loads were applied as
described elsewhere \cite{Mzjctc:09}. This method distributes forces
over selected groups of atoms and compensates them by reactions
applied to other atoms so as to zero the total external force and
torque. Because the forces are applied at different points internal
stress and deformations are introduced that correspond to overall
twisting or stretching. The method was thoroughly verified in Brownian
dynamics simulations of calibrated discrete WLC models
\cite{Mzjctc:09}.

The ranges of forces and torques are selected to comprise the values
used in single molecule manipulation experiments as well as the
corresponding estimates for living cells. It is known that B-DNA
becomes unstable {\em in vitro} with stretching forces beyond 50 pN
\cite{Smith:96,Cluzel:96}. The covalent bonds in long DNA are broken
already with forces beyond 300 pN \cite{Bustamante:00}, and in living
cells DNA is often fragmented during replication in so-called fragile
sites \cite{Letessier:11}. A stretching load of a few tens of
piconewtons can be exerted by a single molecule of RNA polymerase
during transcription \cite{Yin:95,Wang:98}, and forces in the nN range
pull the chromatids during cell division \cite{Nicklas:83}.  The range
of torques that do not destroy B-DNA in single molecule experiments is
from -10 to +35 pNnm \cite{Bryant:03}.  The lower limit is close to
the estimated torsional stress due to natural negative supercoiling in
procaryotes. These data concern the integral stability of long random
sequence DNA. Short stretches of B-DNA can tolerate much stronger
torsional strain. For instance, the DNA twisting observed in complexes
with some bacteriophage repressors \cite{Koudelka:06} corresponds to
torques beyond 100 pNnm. 

The classical MD simulations were carried out by running independent
trajectories in parallel on different processors for identical
conditions. The number of processors varied between 32 and 48.
Trajectories with the lowest loads started from the final states of
free dynamics. The amplitudes of forces and torques were increased
gradually so that simulations with higher values started from the
final states obtained under the preceding lower values. The initial
0.5 ns of every sub-trajectory were discarded, which was sufficient
for re-equilibration.

The AMBER98 forcefield parameters \cite{Cornell:95,Cheatham:99} were
used with the rigid TIP3P water model \cite{Jorgensen:83}. The
electrostatic interactions were treated by the SPME method
\cite{Essmann:95}. To increase the time step, MD simulations were
carried out by the internal coordinate method (ICMD)
\cite{Mzjcc:97,Mzjchp:99}, with the internal DNA mobility limited to
essential degrees of freedom. The rotation of water molecules and DNA
groups including only hydrogen atoms were slowed down by weighting of the
corresponding inertia tensors \cite{Mzjacs:98,Mzjpc:98}. The
double-helical DNA was modeled with all backbone torsions, free bond
angles in the sugar rings, and rigid bases and phosphate groups. The
effect of these constraints is insignificant, as was previously checked
through comparisons with standard Cartesian dynamics
\cite{Mzjacs:98,Mzbj:06}. The time step was 0.01 ps and the DNA structures
were saved every 5 ps. All trajectories were continued to obtain the
sampling corresponding to 164 ns of continuous dynamics, that is 2$^{15}$
points for every value of force (torque).

Additional technical details including preparation of initial states,
treatment of rare events, evaluation of statistical errors, and others
are described elsewhere \cite{Note1}.

\subsection*{Evaluation of elastic parameters}

The DNA elasticity is conveniently characterized by three persistence
lengths (PLs) corresponding to bending, twisting and stretching that
we denote here as $l_b$, $l_t$, and $l_s$, respectively.  These
parameters can be extracted from simulated canonical conformational
ensembles by using the WLC theory that provides linear relationships
of the following form
\beq \label{EDx0}
D_x(L)=\frac{L}{l_x}
\eeq
where $L$ is the DNA length and $x$ stands for $b$, $t$, or $s$.
The WLC deviations $D_x(L)$ are computed from appropriate
canonical averages as
\bea\label{Ewlc2}
D_b(L)&=&-\ln\left(\langle\cos\left[\theta\left(L\right)\right]
        \rangle\right)\nonumber\\
D_t(L)&=&{\bf D}\left[\Phi\left(L\right)\right]\\
D_s(L)&=&\left(\frac{2\pi}{\rm 3.4\ nm}\right)^2
        {\bf D}\left[L\right]\nonumber
\eea
where $\theta(L)$ and $\Phi(L)$ are the angles of bending and twisting,
respectively. The angular brackets denote the canonical averaging and
{\bf D} with square brackets refer to the variance of the variable in
the brackets.  The sampled conformations of the double helix were
analyzed by the program 3DNA \cite{Lu:03c}. Because the elastic
parameters should be preferably estimated by using integral numbers of
helical turns \cite{Mzjpc:09}, only 11 central base pair steps (bps)
were considered (central dodecamers referred to as TA$_6$ and CG$_6$,
respectively).  In the following text, symbols $\theta$, $\Phi$, and
$L$ denote the corresponding parameters of one helical turn.

According to the standard convention \cite{Olson:01}, every base pair
is characterized by a local Cartesian frame, with the xy-plane
parallel to the base pair and z-vectors directed along the DNA. The
bend angle $\theta$ is measured between the z-vectors constructed at
the opposite ends of a helical turn. Earlier it was shown that this
measure of bending is adequate for integral numbers of helical turns
\cite{Mzjpc:09}. The torsional fluctuations were probed by three
alternative methods. The end-to-end twist, $\Phi'$, was evaluated
similarly to the local twist \cite{Lu:97b}, but using the two terminal
reference frames. The cumulated local twist $\Phi''$ is obtained by
summing the local twist at all base-pair steps.  The last angle
$\Phi'''$ is computed similarly by using the base-pair twist with
respect to the optimal helical axis. The fluctuations of the DNA
length were also evaluated by using three alternative methods. The
end-to-end distance, $L'$ was measured directly between the origins of
the terminal reference frames. The contour length $L''$ was measured
by summing the distances between the consecutive frames. The last
value, $L'''$ was obtained by summing the local rise from the 3DNA
output.  These three methods give different average $L$ values and it
is not evident which of them is the best estimate of the macroscopic
DNA length. Therefore, in \Req{Dx0} we used $L$ computed as
$11\cdot0.335$ nm, that is by using the experimental length for one
bps. This can cause a systematic bias in the measured PLs, but does
not affect qualitative trends.

\section*{Results}

\subsection*{Two stretching rigidities of the double helix}

The length of the double helix is usually evaluated by summing the
helical rise along the molecule\cite{Lankas:00,Mzbj:06}. The rise
can be measured with respect to the helical axis (global rise) or
between the base pair frames (local rise). In both cases it is
sensitive to algorithmic differences between the analysis programs
\cite{Lu:99}, and the corresponding $l_s$ values sometimes diverge
very significantly \cite{Mzjpc:09}. To get reliable estimates we
tested several possibilities and three representative techniques
outlined in Methods are compared below. The end-to-end distance,
$L'$, is a direct measure that is adequate in our case because for
very short DNA the length fluctuations are dominated by stretching
\cite{Mathew-Fenn:08b,Mzpre:09}. The second parameter, $L''$, is the
length of the three-dimensional zigzag line through the origins of
the reference frames. By construction, $L'\le L''$ (see \Rtb{zevu}).
The cumulated local rise, $L'''$, was used in the earlier studies
\cite{Lankas:00,Mzjpc:09}. The local rise is one of the orthogonal
projections of the distance between the neighbor frames, therefore,
$L'''\le L''$. A similar value computed with the global rise is not
considered here.

\begin{table}
\caption{\label{Tzevu}
Reference zero stress values of the main parameters discussed in the
text. The DNA length, $L$, the twist, $\Phi$, and the corresponding
PLs, $l_s$ and $l_t$, respectively, were measured by three alternative
methods outlined in Methods. The statistical errors for $L$ and $\Phi$
were about 0.4 deg and 0.04 \AA, respectively. For $l_s$ and $l_t$
the relative errors were about 4\% and 5\%, respectively. }
\begin{ruledtabular}
\begin{tabular}{cccccccccc}
    & \multicolumn{2}{c}{$L$ (\AA)}
    & \multicolumn{2}{c}{$l_s$ (nm)} &
    & \multicolumn{2}{c}{$\Phi$ (deg)}
    & \multicolumn{2}{c}{$l_t$ (nm)}\\
\cline{2-5}
\cline{7-10}
method& TA$_6$ & CG$_6$ & TA$_6$ & CG$_6$ &method& TA$_6$ & CG$_6$ & TA$_6$ & CG$_6$ \\
\hline
 $L'  $ & 34.2 & 35.4 &  78 & 172 & $\Phi'  $ &  -5.4 &  15.8 & 121 & 123\\
 $L'' $ & 38.7 & 38.2 & 230 & 238 & $\Phi'' $ & 349.4 & 372.6 & 102 & 116\\
 $L'''$ & 36.3 & 36.4 & 342 & 394 & $\Phi'''$ & 365.4 & 384.0 & 140 & 122\\
\end{tabular}
\end{ruledtabular}
\end{table}

\Rfg{vsfc0} shows the extension-vs-force plots obtained with the above
three lengths. All three plots are approximately linear, in good
agreement with the harmonic approximation, but the $L'$ value grows
much faster than the other two. Only the vertical positions of the
theoretical straight lines were fitted to the data points while the
slopes were computed independently, which gives an additional check of
self-consistency. The increase of $L''$ is similar to that of
$L'''$ notwithstanding the divergence of their absolute values (see
\Rtb{zevu}), and this increase agrees with the $l_s$ value obtained
from equilibrium fluctuations of $L'''$ rather than $L''$. To explain
these observations, note that the zigzag probed by $L''$ forms a
helical trace that winds around the straight segment measured by $L'$.
\Rfg{vsfc0} suggests that the strokes of the zigzag can be considered
inextensible, and the end-to-end distance grows mainly due to
flattening of angles. The local helical parameters are obtained
by decomposing each stroke of the zigzag into rise, shift and slide
\cite{Lu:97b}. All three of them contribute to the fluctuations of
$L''$, however, only the rise is affected by the applied force
because the other two correspond to displacements nearly orthogonal
to the force. This explains why the growth of $L''$ is better
described by $l_s$ obtained from $L'''$.

\begin{figure}[ht]
\centerline{\includegraphics[width=8cm]{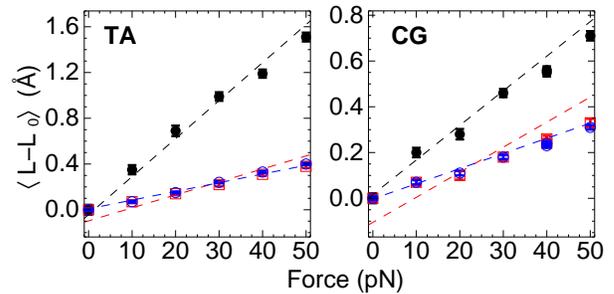}}
\caption{ \label{Fvsfc0} Color online.
DNA extension under steady stretching load. The molecule length was
measured by three different methods outlined in the text. The data for
$L'$, $L''$ and $L'''$ are shown by black dots, open red squares, and
open blue circles, respectively. The error bars are small and they
merge with symbols.  The theoretical harmonic dependences (colored
dashed lines) were plotted for the zero-stress values of stretching PL
presented in \Rtb{zevu}, with the vertical shifts fitted to the data
points. The left and right panels exhibit the results for TA$_6$ and
CG$_6$, respectively.
}\end{figure}%=====================================================

According to \Rfg{vsfc0} the double helix is characterized by two
qualitatively different stretching rigidities. Parameter $l^{'}_s$
corresponding to fluctuations of $L'$ can be measured experimentally.
In the experimental literature the stretching stiffness is
conventionally characterized by the modulus $Y_f$ related to $l_s$
as
$$
l_s=\frac{Y_f}{kT}\left(\frac{\rm 3.4nm}{2\pi}\right)^2.
$$
The experimental estimates of $Y_f$ are around 1100 pN
\cite{Smith:96,Wang:97b,Wenner:02}, which corresponds to $l_s$=78 nm,
in reasonable agreement with $l^{'}_s$ computed from MD data (see
\Rtb{zevu}). This conclusion is corroborated by \Rfg{vstq0} that shows
how the measured DNA length changes with forced twisting.  According
to experiments \cite{Gore:06b,Lionnet:06} small twisting should cause
extension of the double helix. It is seen that the end-to-end length
$L'$ indeed grows with small twisting in quantitative agreement with
the experimental estimate.

\begin{figure}[ht]
\centerline{\includegraphics[width=8cm]{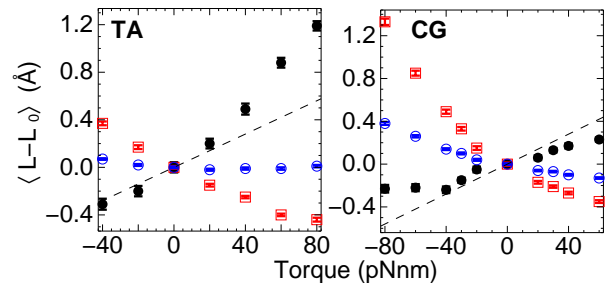}}
\caption{ \label{Fvstq0} Color online.
Variations of the DNA length under steady twisting torques. The three
plots in each panel correspond to alternative definitions of DNA
length explained in the text. The notations are same as in
\Rfg{vsfc0}. The dotted traces show the expected dependences for a
harmonic twist-stretch coupling with parameters measured
experimentally \cite{Gore:06b}.
}\end{figure}%=====================================================

In contrast, the stretching rigidity characterized by parameter
$l^{'''}_s$ is similar for both sequences, but significantly larger
than the experimental estimate (see \Rtb{zevu}). Thermal
fluctuations of the local rise involve perturbations of base-pair
stacking, therefore, $l^{'''}_s$ specifically characterizes the
strength of stacking interactions.  However, this stretching
rigidity is not probed in experiments. \Rfg{vstq0} reveals that the
lengths measured by parameters $L''$ and $L'''$ both decrease with
twisting, in qualitative divergence from $L'$ and experimental
observations \cite{Gore:06b,Lionnet:06}.

\begin{figure}[ht]
\centerline{\includegraphics[width=8cm]{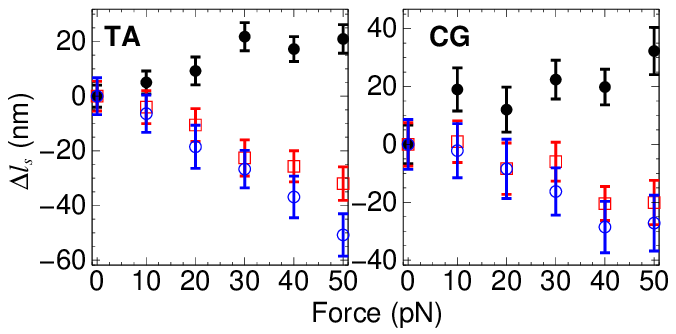}}
\caption{ \label{Fvsfc1} Color online.
Effect of tension upon the stretching rigidity of DNA. The stretching
PL was evaluated from the length fluctuations measured by three
alternative methods outlined in the text. The reference zero stress
values are presented in \Rtb{zevu}.  The notations correspond to
\Rfg{vsfc0}, namely, the data for $l^{'}_s$, $l^{''}_s$, and
$l^{'''}_s$ are shown by black dots, open red squares, and open blue
circles, respectively.
}\end{figure}%=====================================================

The stretching rigidity does not remain constant with forced
stretching and twisting. \Rfg{vsfc1} reveals that in stretched DNA,
$l^{'}_s$ and $l^{'''}_s$ deviate in opposite senses. The $l^{'}_s$
value corresponding to experimental measurements grows. Therefore, the
molecule should gradually become stiffer until the stretching force
approaches the limit of about 70 pN where the B-DNA is known to
loose stability \cite{Strick:00a}. The stiffening agrees with the
deviations of black points in \Rfg{vsfc0} from the linear plots
corresponding to the harmonic approximation. Mechanistically, the
growth of $l^{'}_s$ can be rationalized by noting that, with the zigzag
angles flattened, the end-to-end distance $L'$ approaches the zigzag
length $L''$.  Since $L'$ can never exceed $L''$, the fluctuations of
$L'$ should decrease, that is, $l^{'}_s$ grows approaching $l^{''}_s$.
The simultaneous decrease of $l^{'''}_s$ reflects gradual weakening of
base-pair stacking. Twisting also increases the stretching rigidity
(see \Rfg{vstq1}). However, untwisting of TA$_6$ changes $l^{'}_s$ only
slightly, suggesting that it passes through a minimum with torque
$\tau$=-20 pNnm. Interestingly, the value of $l^{'}_s$ reached
with untwisting of CG$_6$ is similar to that of TA$_6$.

\begin{figure}[ht]
\centerline{\includegraphics[width=8cm]{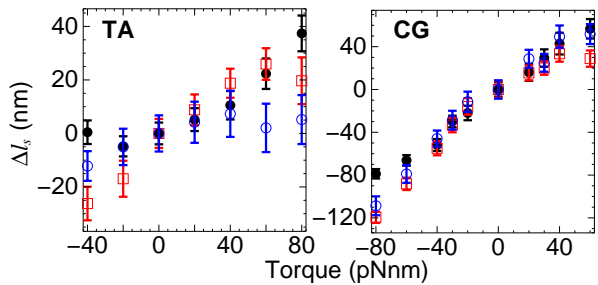}}
\caption{ \label{Fvstq1} Color online.
Effect of forced twisting upon the stretching rigidity of DNA. The
three plots in each panel correspond to alternative definitions of DNA
length explained in the text. The notations are same as in
\rfg{vsfc1}.
}\end{figure}%=====================================================

\subsection*{Torsional rigidity}

In the previous report \cite{Mzprl:10} the torsional rigidity was
evaluated by using the twist angle $\Phi'''$ (see Methods). This
parameter depends upon the construction of an optimal straight
helical axis, which can add a spurious noise due to bending
deformations of the double helix. For verification, here the
torsional dynamics are analyzed by three alternative methods
including the earlier one. The end-to-end twist $\Phi'$ is most
appropriate for comparisons with experiment because it closely
corresponds to that measured in experiments with long DNA. The
cumulated local twist $\Phi''$ represents another reasonable
alternative and it was added as an additional check.

\begin{figure}[ht]
\centerline{\includegraphics[width=8cm]{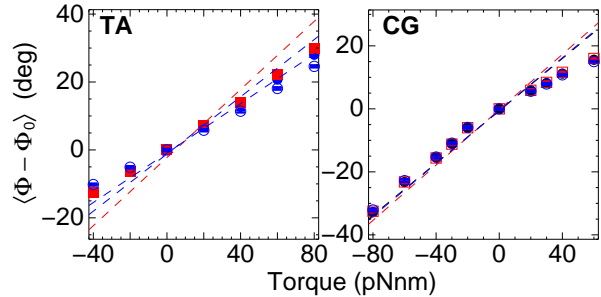}}
\caption{ \label{Fvstq2} Color online.
DNA twisting under steady torques. The twist angle was measured by
three different methods outlined in the text. The data for $\Phi'$,
$\Phi''$, and $\Phi'''$ are shown by black dots, open red squares, and
open blue circles, respectively. The error bars are small and they
merge with symbols. The theoretical harmonic dependences (colored
dashed lines) were plotted for the corresponding zero-stress values of
the torsional PL presented in \Rtb{zevu}, with the vertical shifts
fitted to the data points. The left and right panels exhibit the
results for TA$_6$ and CG$_6$, respectively.
}\end{figure}%=====================================================

\begin{figure}[ht]
\centerline{\includegraphics[width=8cm]{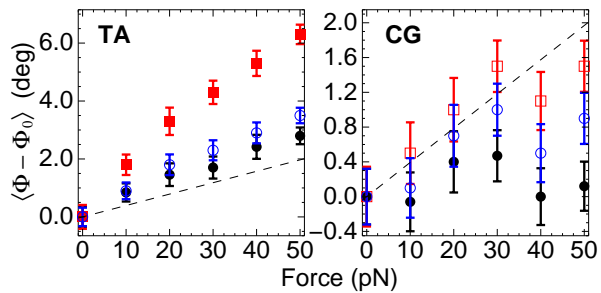}}
\caption{ \label{Fvsfc2} Color online.
Effect of stretching upon DNA twisting. The three plots in each panel
correspond to alternative measurements of DNA twisting as explained in
the text. The notations are same as in \Rfg{vstq2}.  The dotted traces
show the expected dependences for a harmonic twist-stretch coupling
with parameters measured experimentally for random-sequence DNA
\cite{Gore:06b}.
}\end{figure}%=====================================================

The external torque changes DNA twisting as shown in \Rfg{vstq2}. In
contrast to stretching, the three alternative measures of angle $\Phi$
give very similar results in spite of the divergence of the reference
zero-stress values (see \Rtb{zevu}). Similarly to \Rfg{vsfc0}, only
the vertical positions of the theoretical straight lines were fitted
to the data points, with the slopes computed independently. This
additionally checks the self-consistency and one may note that the
deviations from the harmonic law are smaller for $\Phi'$ and $\Phi'''$
than for $\Phi''$. Earlier single-molecule experiments revealed that
DNA overwinds when stretched \cite{Gore:06b,Lionnet:06}.  This effect
is well reproduced with any of the three methods (see \Rfg{vsfc2})
sometimes with good quantitative agreement. The dashed lines in
\Rfg{vsfc2} represent the experimental dependence for small forces
below 30 pN  \cite{Gore:06b}. With stronger extension the twist should
start to fall. For TA$_6$ this experimental observation is not
reproduced, but for CG$_6$ a transition from an ascending trend to an
irregular plateau is indeed observed at about 30 pN.  This irregular
dependence is not due to errors or hidden statistical noise. For
verification, we reduced the force from 50 to 40 pN, repeated the MD
simulations, then raised the force back to 50 pN, and carried out one
more run. The results of this back and forth test were within the
error limits shown in \Rfg{vsfc2}.

\begin{figure}[ht]
\centerline{\includegraphics[width=8cm]{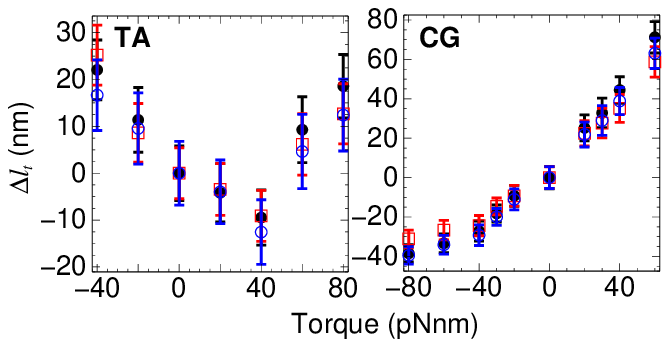}}
\caption{ \label{Fvstq3} Color online.
Effect of forced twisting upon the torsional rigidity of DNA. The
torsional PL was evaluated from the twist fluctuations measured by
three alternative methods outlined in the text. The reference zero
stress values are presented in \Rtb{zevu}.  The notations correspond to
\Rfg{vstq2}, namely, the data for $l^{'}_t$, $l^{''}_t$, and
$l^{'''}_t$ are shown by black dots, open red squares, and open blue
circles, respectively.
}\end{figure}%=====================================================

The measured torsional rigidity changes with forced twisting as shown
in \Rfg{vstq3}. The three alternative measures of twist yield very
similar results all showing strong variations of $l_t$, with a
remarkable qualitative difference between the two sequences. These
rigidity profiles agree with the non-linear features of the
$\Phi(\tau)$ plots in \Rfg{vstq2}. Indeed, for CG$_6$ they are concave
and for TA$_6$ the harmonic law corresponding to the zero-stress
rigidity overestimates the twisting of both signs.  The twisting
rigidity of CG$_6$ grows steadily in the whole range of torques
tested. In contrast, for TA$_6$ an opposite trend is observed under
small torques, but $l_t$ passes via a minimum under positive torques.
A qualitatively similar behavior was experimentally observed for one
natural DNA sequence \cite{Song:90b,Naimushin:94}.

The growth of rigidity with torques of both signs agrees with simple
physical intuition for a twist energy profile resembling a
flat-bottomed basin with vertical walls. In this case the system
cannot go very far even with strong energy fluctuations. The range of
torques applied to CG$_6$ was extended to check the existence of a
minimum under negative torques. It is seen, however, that the minimum
is not reached although the decrease of $l_t$ becomes less steep with
untwisting. This behavior indicates that anomalously frequent strong
untwisting fluctuations should occur in GC-alternating DNA under
normal temperature.

\begin{figure}[ht]
\centerline{\includegraphics[width=8cm]{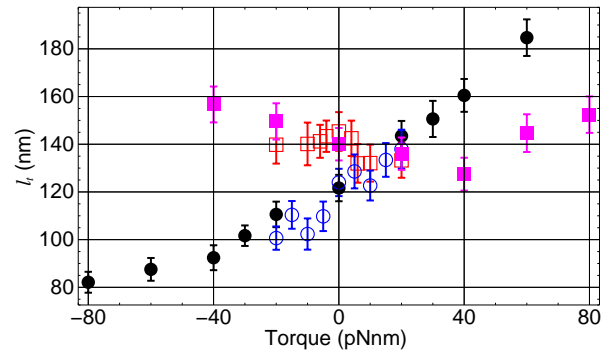}}
\caption{ \label{Fvstq4} Color online.
Twisting rigidity of DNA under small and large torques. The results of
the earlier report \cite{Mzprl:10} (open symbols) are compared with
those of the present study (filled symbols). Squares and circles show
the data data for TA$_6$ and CG$_6$, respectively. The DNA twisting
was evaluated by using angle $\Phi'''$.
}\end{figure}%=====================================================

The results in \Rfg{vstq3} confirm and corroborate the conclusions of the
previous report where smaller torques were considered \cite{Mzprl:10}.
The earlier data are compared with those of the present study in
\Rfg{vstq4}. It is seen that the two series of simulations are
consistent in spite of the differences in protocols. Each open
circle and open square in \Rfg{vstq4} correspond to a single
continuous trajectory, therefore, this figure confirms ergodicity
and validates the much faster protocol introduced here. The new plots
also look less noisy, which can be attributed to the absence of slow
non-canonical $\alpha/\gamma$ dynamics. In the previous calculations,
such transitions occurred almost exclusively in terminal bps
\cite{Mzjctc:09,Mzprl:10}, nevertheless, they affected the
middle fragments allosterically and contaminated the results.

\begin{figure}[ht]
\centerline{\includegraphics[width=8cm]{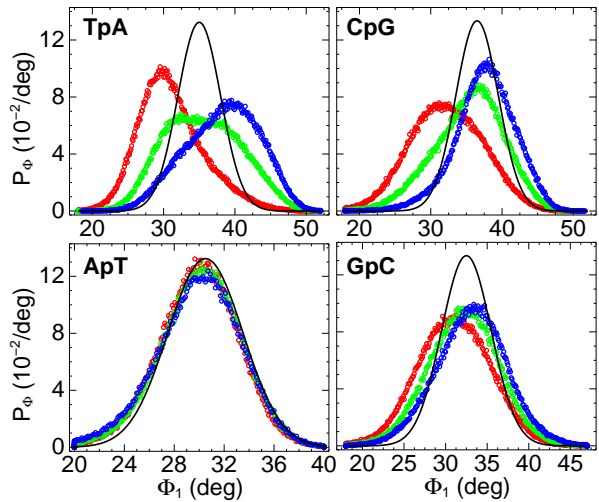}}
\caption{ \label{Fmdpbd1} Color online.
The normalized probability densities of twisting fluctuations in the
four types of bps represented in TA$_6$ and CG$_6$ fragments. Open
circles show the computed distributions for strong negative (left,
red), medium (middle, green), and strong positive (right, blue)
values of external torque, respectively. The corresponding torque
values were $\tau$= -40, +40, and +80 pNnm for TA$_6$ and $\tau$=
-40, 0, and +40 pNnm for CG$_6$. The black solid lines show the
analytical Gaussian distributions for $l_t$ measured under the
middle torque values. These curves are placed to match the maxima of
the corresponding computed distributions.
}\end{figure}%=====================================================

The strong torsional anharmonicity is not seen in the shapes of
the probability distributions of twisting fluctuations of the whole
fragment. These distributions remain nearly Gaussian, with the widths
changing in agreement with \Rfg{vstq3} (see Refs.
\onlinecite{Mzprl:10,Note1}). In contrast, the pattern of single-step
twist fluctuations qualitatively explains the effect revealed in
\Rfg{vstq3} and \rfg{vstq4}. As seen in \Rfg{mdpbd1}, with a notable
exception of the adenine-phosphate-thymine steps (ApT), these
distributions strongly differ from Gaussians predicted for harmonic WLC
model with the measured $l_t$ values. Surprisingly, for TpA and CpG
steps these shapes qualitatively change with twisting. With negative
torques the distributions in the upper two panels are strongly
positively skewed, but they gradually become negatively skewed as the
torque changes the sign. The same is true for the GpC distributions
although in this case the effect is much smaller. Some of the TpA and
CpG distributions exhibit clear humps suggesting that the twisting in
these steps is best described by double-well potentials with low
transition barriers.

The CpG and GpC distributions in \Rfg{mdpbd1} behave similarly, that
is, they become wider with untwisting in qualitative agreement with
the $l_t(\tau)$ plots for CG$_6$ in \Rfg{vstq4}. In contrast, in
TA$_6$ the two alternating dinucleotide steps behave differently. The width
of the ApT step distributions changes monotonously in the whole range
of torques probed, that is, the minimum of $l_t$ at 40 pNnm in
\Rfg{vstq4} is exclusively due to TpA steps. The ApT distributions
also exhibit a striking feature. The centers of all plots are
shifted in agreement with the sign of the applied torque, however,
the magnitude of the shift is small compared to the change in the
distribution width. As a result, the probabilities of strong
untwisting fluctuations are higher with $\tau$=+80 pNnm than with
$\tau$=-40 pNnm. The effect is small, but statistically significant
(see also Ref. \onlinecite{Note1}). This feature is counterintuitive
because it cannot be reproduced with the WLC model.

\begin{figure}[ht]
\centerline{\includegraphics[width=8cm]{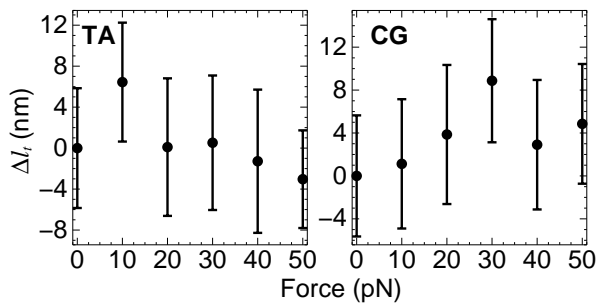}}
\caption{ \label{Fvsfc3}
Effect of stretching forces upon the torsional rigidity of DNA. The
measured values of $l^{'}_t$ are shown for TA$_6$ and CG$_6$ on the
left and right panels, respectively.
}\end{figure}%=====================================================

In contrast to twisting, small stretching has virtually no effect
upon the torsional rigidity of DNA. The corresponding data are shown
in \Rfg{vsfc3}. For clarity, only the $l^{'}_t$ values are shown.
The variations are small and rarely exceed the statistical errors.
When $l_t$ is measured by using magnetic tweezers the common
stretching load is smaller than 20 pN \cite{Bryant:03} and the data
in \Rfg{vsfc3} indicate that it can noticeably affect the results
only due to mechanisms that are not reproduced in the present DNA
model.

\subsection*{Bending rigidity}

In long DNA, stretching naturally flattens bends, whereas twisting
causes looping and supercoiling, that is, increases bending in some DNA
stretches. These effects are strong; the accompanying changes in the
bending rigidity are hardly measurable experimentally and this
possibility usually is not considered. The atom-level modeling is the
only currently available method that can check whether or not the
bending rigidity of DNA in principle can be affected by the twisting
and/or tensional stress. The results of the first such tests are
shown in \Rfg{bpl}.

\begin{figure}[ht]
\centerline{\includegraphics[width=8cm]{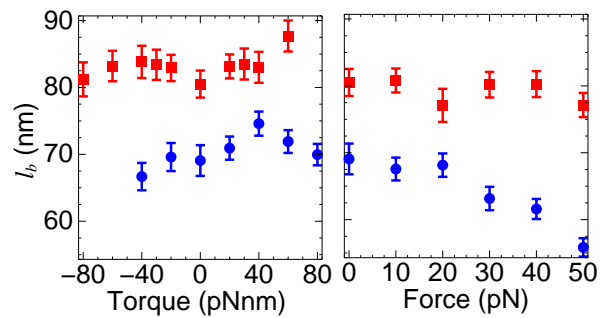}}
\caption{ \label{Fbpl} Color online.
Variation of the apparent bending PL with forced stretching and
twisting. The results for CG$_6$ and TA$_6$ are shown by red squares
and blue circles, respectively.
}\end{figure}%=====================================================

The measured bending PL of CG$_6$ exhibits only small variations with
both twisting and stretching. In contrast, for TA$_6$ these variations
significantly exceed statistical uncertainty and reveal interesting
trends. Notably, the right panel of \Rfg{bpl} reveals that bending in
TA$_6$ increases with stretching, which is opposite to the expected
flattening effect. At the same time, the $l_b(\tau)$ dependence in the
left panel of \Rfg{bpl} passes through a maximum at 40 pNnm,
that is exactly where the torsional PL reaches the local minimum in
\Rfg{vstq3}. A closer look reveals that these trends are accompanied
by subtle qualitative changes in the bending dynamics. By using the
base pair coordinate frames provided by the program 3DNA \cite{Lu:03c}
one can conveniently characterize the bend direction as follows.
Consider two coordinate frames constructed at the first and the last
base pairs, respectively. According to the standard convention
\cite{Olson:01}, the two xy-planes dissect the double helix
approximately parallel to the base pair planes. The corresponding two
z-vectors approximate the local directions of the helical axis. If the
z-vectors are not parallel we can construct the orthogonal projection
of the second z-vector upon the first xy-plane. The spherical azimuth
angle $\varphi$ is measured between the projected z-vector and the
x-vector of the projection plane. With the x-vector corresponding to
the standard convention \cite{Olson:01}, the value of $\varphi$ is
close to zero when the molecule is bent towards the minor groove in
the middle of the helical turn. A few representative distributions of
angle $\varphi$ are shown in \Rfg{mdpbd3}.

\begin{figure}[ht]
\centerline{\includegraphics[width=8cm]{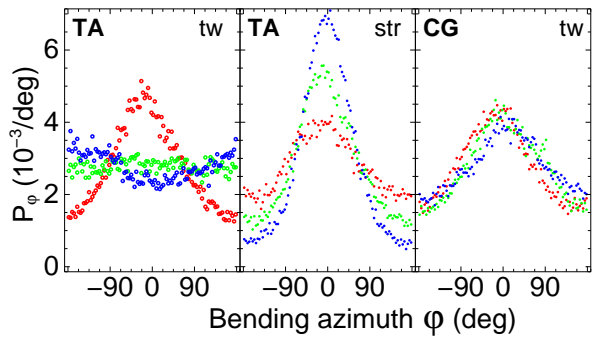}}
\caption{ \label{Fmdpbd3} Color online.
Effects of stretching and twisting upon the anisotropy of bending.
The normalized probability distributions for the bending azimuth angle
$\varphi$ are shown for TA$_6$ twisting (left panel) and stretching
(middle panel), and for CG$_6$ twisting (right panel). The red, blue,
and green dots correspond to torques $\tau$= -40, +40, and +80
pNnm (left panel), forces of 0, 20, and 50 pN (middle panel),
and  torques $\tau$= -80, 0, and +60 pNnm (right panel).
}\end{figure}%=====================================================

The red distribution in the middle panel indicates that the unstressed
TA$_6$ fragment prefers to bend towards the minor groove. The origin
of this anisotropy should be studied additionally because it is
probably inherent in the overall dynamics rather than caused by local
end effects or construction of the coordinate frames. Here we use it
just as an indicator. It is seen that the original anisotropy increases
with both stretching and unwinding, but positive torques reduce it. As
a result, with $\tau$=40 pNnm the azimuth distribution becomes
even, and with further twisting the anisotropy of an opposite sign
appears (left panel). This behavior is in remarkable contrast to that
of CG$_6$. For CG$_6$ the bending is also preferable towards the
minor groove, but twisting only causes rotation of this direction in
agreement with the relative orientation of the minor groove in the
middle of the fragment (right panel).

Comparison of \Rfg{mdpbd3} with \Rfg{vstq3} and \rfg{mdpbd1} suggests
that $\tau$=40 pNnm corresponds to a transition state between
two qualitatively different dynamic patterns and that this behavior is
attributable to the specific properties of TpA steps. The results in
\Rfg{vstq3} and \rfg{mdpbd1} can be readily rationalized and
qualitatively reproduced in an appropriate coarse-grained model with
local twisting described by a double-well potential. Global bending of
the double helix results from local deviations of bps geometry
described by parameters Roll, Tilt, Slide, and Shift \cite{Olson:01}.
Analysis shows that, in TpA steps, all of them are affected by
twisting.  Roll and Slide change more than other, with Slide
exhibiting a bi-modal pattern of fluctuations \cite{Note1}. These data
demonstrate that a mechanical link between twisting and bending is an
inherent property of TpA dinucleotides, therefore, we can qualitatively
explain the results in the left panel of \Rfg{mdpbd3}. The quantitative
relation is much more difficult to establish because global bending
results from a complex summation of local motions over the whole
fragment, including correlations and helical rotation. The
correspondence of the transition states for twisting and bending in
TA$_6$ may be a coincidence, nevertheless, the results in \Rfg{bpl} and
\rfg{mdpbd3} evidence that twisting and stretching can produce
unexpected sequence dependent effects upon the bending dynamics in DNA.
The stress response is complex and it cannot be reduced to an altered
bending rigidity of the underlying WLC model.

\subsection*{Discussion}

A very good agreement of experiments on polymer DNA with the WLC model
\cite{Bustamante:94,Vologodskii:94a,Vologodskii:97,Wang:97b,Moroz:97,Bouchiat:99}
have led to an exaggerated belief that the harmonic approximation is
sufficient for describing all essential properties of the DNA double
helix. In fact, these remarkable results cannot be considered as
evidences of harmonicity because the additive ladder construction of
the double helix effectively hides local heterogeneity and
anharmonicity. Due to this additivity, and the central limiting
theorem of the probability theory, various experimental data converge
to the WLC model regardless of the local DNA properties, with only a
few concatenated bps being sufficient for the apparent statistical
equivalence with a harmonic elastic rod \cite{Mzprl:10}. This effect
shadows the true mechanical properties of the DNA double helix which
remain elusive.

The present study evidences that, under normal temperature, the local
DNA elasticity is strongly anharmonic, in agreement with the early
hypotheses \cite{Shibata:84} and some experimental data
\cite{Song:90b,Selvin:92,Naimushin:94}. The results of computations
using empirical forcefields certainly require further verification.
New experimental approaches need to be developed for this purpose
because currently available methods can probe only the average elastic
parameters of long molecules.

The computed values of all elastic parameters reasonably agree with the
data for polymer DNA obtained by different experimental methods. The
earlier controversy concerning the stretching (Young's) modulus
\cite{Mzjpc:09} is clarified here by comparing different procedures for
measuring the length of the double helix. The experimental bending
rigidity is characterized by $l_b\approx50$ nm \cite{Hagerman:88}. The
measured $l_t$ values vary between 36 and 109 nm depending upon
specific methods and conditions \cite{Fujimoto:06}.  The stretching
PL is about 80 nm
\cite{Smith:96,Wang:97b,Wenner:02}. MD simulations give somewhat larger
values, that is, the DNA stiffness is slightly overestimated.
\cite{Lankas:00,Mzbj:06} This discrepancy is not large and it can be
attributed to a combination of factors like inexact correspondence
between the microscopic geometric parameters and experimental
observables, the neutralizing salt conditions in MD, and the small size
of the modeled fragments predictably leading to strong sequence and end
effects. As shown here, MD also quantitatively reproduce the reciprocal
coupling between twisting and stretching revealed in recent magnetic
tweezer experiments \cite{Gore:06b,Lionnet:06}. The overall agreement
is quite remarkable because none of the MD forcefield parameters was
adjusted to fit the computed DNA elasticity to experiment.  One may
reasonably hope, therefore, that the detailed microscopic picture
provided by simulations captures the qualitative physical trends
dictated by the atom-level mechanics of the double helix.

Our results indicate that the most significant anharmonicity is
inherent in the torsional DNA deformations, which is attributable to
the special character of stacking interactions. The twisting occurs due
to sliding within the stacks; this motion is essentially barrierless
and its amplitude significantly exceeds the zone where the harmonic
approximation is valid. Even small twisting torques can cause
significant changes in elastic parameters. The qualitative
difference in the stress response of the torsional rigidity of AT$_6$
and CG$_6$ indicates that this property is strongly sequence-dependent.
Opposite local trends can mutually cancel out, which makes difficult
detection of anharmonic effects in long DNA.  There are a few reports
in the literature where relevant experimental data qualitatively differ
from predictions of harmonic models.  This occurred with some natural
plasmid DNA \cite{Naimushin:94} and also with synthetic alternating
sequences \cite{Wells:77}. The latter were recently found anomalous as
regards the sequence-dependent bending rigidity \cite{Geggier:10}.
These earlier results require additional investigations.

The mechanical strain is an ubiquitous attribute of living DNA and a
key factor in genome packaging and regulation.  The common magnitudes
of natural forces and torques are quite large, therefore, a wide
spectrum of non-linear structural responses should be anticipated, with
elastic deformations at one end of the scale, and local melting at the
other end. A few anharmonic effects revealed here have some interesting
implications for gene regulation mechanisms.  According to \Rfg{vstq4},
with the helical twist slightly shifted from the equilibrium value the
sequence dependence of the DNA elasticity can be significantly changed
and enhanced. The measured torsional stiffnesses are similar without
applied torque, but diverge with both twisting and untwisting. For
other sequences, similar behavior can be anticipated for bending and
stretching. The deformability of DNA is long considered as a possible
governing factor in the sequence-specific site recognition
\cite{Hogan:87}, but this mechanism requires strong sequence dependence
of local elastic parameters compared to that observed in experiments
with long free DNA \cite{Fujimoto:90}. As we see the properties of the
relaxed DNA cannot be simply transferred to supercoiled and/or protein
bound DNA states. Additional studies are necessary to check if the
elastic properties of the specific binding sites are sensitive to
external stress.

Unexpectedly, we found that the torsional rigidity of AT$_6$ passes via
a minimum under moderate positive twisting torques. This feature is
probably due to a bimodal character of twist fluctuations in the TpA
steps (\Rfg{mdpbd1}). The average twist of AT$_6$ with $\tau$=40
pNnm actually is very close to the experimental value in
solution \cite{Rhodes:81,Strauss:81} because in free AMBER simulations
the DNA structures are somewhat underwound \cite{Cheatham:99}. In this
state the TpA steps exhibit a distribution of twist fluctuations
corresponding to a saddle point between two domains of attraction
(\Rfg{mdpbd1}). This point also coincides with the maximum in the
measured bending PL accompanied by inversion of the local bending
anisotropy.

Earlier it was suggested that the TpA steps can adopt at least two
distinct conformational states. Depending on the sequence context, there
is always a temperature range where the TpA steps exhibit slow
conformational transitions with relaxation times beyond the nanosecond
time range \cite{McAteer:95}. These slow motions should involve
extended DNA stretches, that is, these are global transitions
accompanied by switching in the TpA steps. The same local switching is
probably responsible for the unusual effects observed here. The
exceptional properties of the TpA steps are long-known in molecular
biology \cite{Travers:87}.  These steps are found in both narrowings
and widenings of the minor B-DNA groove \cite{Yoon:88,Quintana:92}.
Periodically spaced TpA steps is the most statistically significant
feature of DNA sequences that provide optimal DNA wrapping around
nucleosome particles \cite{Takasuka:10}. Switching of local bending
anisotropy in response to variable torsional stress may play some role
in the control of DNA wrapping and unwrapping. Future studies will
show whether or not these processes are related with the unusual
microscopic dynamics revealed in our computations.

According to \Rfg{mdpbd1} the strong variation of the twisting rigidity
of CG$_6$ is mainly due to CpG steps. They exhibit anomalously high
probability of negative twist fluctuations with torques around zero.
The CpG steps are found in a number of known protein binding sites, but
their most important biological role is related with C5-cytosine
methylation and epigenetic regulation mechanisms \cite{Robertson:05}.
The recognition of CpG sites is a complex multi-facet process because
they exist in three methylation states with distinct functions
and because specific binding, methylation and demethylation can occur
on both free and nucleosome bound DNA
\cite{Arita:08,Metivier:08,Ho:08,Chodavarapu:10}. Interestingly,
methylation of free DNA strongly depends upon supercoiling, with the
superhelical density acting smoothly in a dose-dependent manner
\cite{Bestor:87}. The corresponding catalytic mechanism requires
cytosine flipping from the DNA stack into a protein pocket
\cite{Klimasauskas:94}. The low energy pathway of this flipping
transition may require a strong twisting fluctuation of the CpG step,
which would explain the effect of the torsional strain
\cite{Bestor:87}.

The above specific examples suggests a more general hypothesis
concerning the possible role of strong DNA fluctuations in gene
regulation, with the non-linear elasticity as the governing factor.
There are many long-known and well-documented processes {\em in vivo}
where strongly deformed conformations are involved instead of canonical
B-DNA. Deformed DNA conformations are ubiquitous in X-ray structures of
protein-DNA complexes, so that one may wonder why there is no
evolutionary pressure towards proteins that can recognize relaxed
B-DNA? It was shown that the activity of promoters regulated via
strongly deformed DNA states can be increased by mutations that reduce
the deformation energy \cite{Chow:91,Li:99b}, but these mutations are
not selected {\em in vivo}. It is possible that the prevalence of large
DNA deformations is not a trivial consequence of its flexibility, but a
necessity of regulatory mechanisms that involve mechanical stress. The
larger the deformation - the lower its probability and the population
of such state. However, these low probabilities can strongly change in
response to small regulatory impulses, in contrast to populations of
low energy states. The non-linear elastic effects should play an
important role in such regulation because they can greatly amplify the
input signal and also make possible complex responses like coupling of
the amplitude and the anisotropy of local bending to the torsional
stress as in the TA$_6$ fragment studies here. Similar ideas were
discussed in the earlier literature. This hypothesis is complementary
to the view of DNA as an allosteric protein cofactor \cite{Lefstin:98}
used to explain the smooth modulation of gene activity during cell
development \cite{Meijsing:09}. The effects of mechanical strain upon
the probabilities of strong fluctuations in DNA represent significant
interest and require further studies. New insights in this direction
can be obtained by using MD simulations of DNA in steady stress
conditions \cite{Mzjctc:09} and this work is continued.

\begin{acknowledgments}
The author is grateful to Mickey Schurr and Andrew Travers for useful
discussions and valuable comments to the original manuscript.
\end{acknowledgments}%............................................

\section{Appendix}
\setcounter{figure}{0}
 \captionsetup{labelformat=empty,labelsep= none,
  justification=centerlast,width=.9\columnwidth,aboveskip=5pt}

\section*{Simulation protocols}

Tetradecamer DNA fragments were modeled with AT-alternating
(ATATATATATATAT) and GC-alternating (GCGCGCGCGCGCGC) sequences. The
choice of the fragment length and sequences is consistent with the
recent computations \cite{Mzjpc:08,Mzjctc:09,Mzjpc:09} and it was
dictated by the following considerations. An integral number of
helical turns is preferable for evaluation of the elastic parameters
of DNA and one helical turn is optimal  because of the rapid growth of
the principal relaxation times with the chain length \cite{Mzjpc:09}.
These molecules are homopolymers of ApT and GpC dinucleotides,
therefore, they cannot have distinguished asymmetric structures like
static bends.  True homopolymer DNA duplexes have special properties
and, in free MD with the AMBER forcefield, these structures deviate
from the canonical B-DNA stronger than AT- and GC-alternating
sequences \cite{Mzjctc:05}.

The classical MD simulations were carried out by running independent
trajectories in parallel on different processors for identical
conditions. The number of processors varied between 32 and 48. The
starting states were prepared as follows.  The solute in the canonical
B-DNA conformation \cite{Arnott:72} was immersed in a 6.2-nm cubic
cell with a high water density of 1.04. The box was neutralized by
placing Na$^+$ ions at random water positions at least 5 \AA\ from the
solute. The system was energy minimized and dynamics were initiated
with the Maxwell distribution of generalized momenta at low
temperature. The system was next slowly heated to 293 K and
equilibrated during 1.0 ns. After that the water density was adjusted
to 0.997 by removing the necessary number of water molecules selected
randomly at least 5 \AA\ from DNA and ions, and the simulations were
continued with NVT ensemble conditions. Independent starting states
for parallel trajectories were prepared by redistributing the
counterions around DNA and re-equilibrating the system with different
random sets of initial momenta. The statistical independence was
verified as explained further below. Steady stress loads were applied
as described elsewhere \cite{Mzjctc:09}. Trajectories with the lowest
loads started from the final states of free dynamics.  The amplitudes
of forces and torques were increased gradually so that simulations
with higher values started from the final states obtained under the
preceding lower values. The initial 0.5 ns of every sub-trajectory
were discarded, which was sufficient for re-equilibration.

The AMBER98 forcefield parameters \cite{Cornell:95,Cheatham:99} were
used with the rigid TIP3P water model \cite{Jorgensen:83}. The
electrostatic interactions were treated by the SPME method
\cite{Essmann:95}, with the common values of Ewald parameters, that is
9 \AA\ truncation for the real space sum and $\beta\approx 0.35$. The
temperature was maintained by the Berendsen algorithm
\cite{Berendsen:84} applied separately to solute and solvent with a
relaxation time of 10 ps.  To increase the time step, MD simulations
were carried out by the internal coordinate method (ICMD),
\cite{Mzjcc:97,Mzjchp:99} with the internal DNA mobility limited to
essential degrees of freedom and rotation of water molecules and
internal DNA groups including only hydrogen atoms slowed down by
weighting of the corresponding inertia tensors.
\cite{Mzjacs:98,Mzjpc:98} The double-helical DNA was modeled with all
backbone torsions, free bond angles in the sugar rings, and rigid
bases and phosphate groups. The effect of these constraints is
insignificant, as was previously checked through comparisons with
standard Cartesian dynamics \cite{Mzjacs:98,Mzbj:06}. The time step
was 0.01 ps and the DNA structures were saved every 5 ps. All
trajectories were continued to obtain the sampling corresponding to
164 ns of continuous dynamics, that is 2$^{15}$ points for every value
of force (torque). Statistical convergence and errors were evaluated
by the method of block averages (see further below).

The terminal base pairs open rather frequently during nanosecond time
scale MD, which significantly perturbs the flanking DNA structure.
Because this dynamics cannot be averaged during the accessible
duration of MD trajectories, we blocked it by applying non-perturbing
upper distance restraints as explained elsewhere \cite{Mzjctc:09}. A
similar approach was used to confine the backbone dynamics to the
canonical B-DNA zones. Rare non-canonical
$\gamma^{g+}\rightarrow\gamma^t$ switches accompanied by
$\alpha^{g-}/\alpha^{g+}$ and $\beta^{t}/\beta^{g+}$ dynamics
complicate statistical analysis of MD trajectories \cite{Dixit:05a}.
The population of non-canonical conformers is overestimated with all
versions of the AMBER94 forcefield including {\em parmbsc0}
\cite{Perez:07a} most used in recent years \cite{Lankas:10}. With the
{\em parm98} modifications \cite{Cheatham:99} designed to increase the
average DNA twist, this problem seems to be less severe
\cite{Mzjctc:09}. However, even with a hypothetical ideal forcefield,
such transitions should have been blocked because they are too rare
for accurate averaging. It appears that the following flat-bottom
restraint upon $\gamma$ torsions solves the problem
$$
U_\gamma=\left\lbrace\begin{array}{cr}
                 A_\gamma(\gamma-2\pi/3)^2,& \gamma\ge2\pi/3\\
                 0,                & -\pi<\gamma<2\pi/3\\
                 \end{array}\right.
$$
with $A_\gamma\approx$30 kcal/mol. This noninvasive approach does not
perturb the dynamics in the canonical zones, which makes possible
comparison with the previous long-time simulations of similar DNA
fragments.

\begin{figure}[ht]
\centerline{\includegraphics[width=8cm]{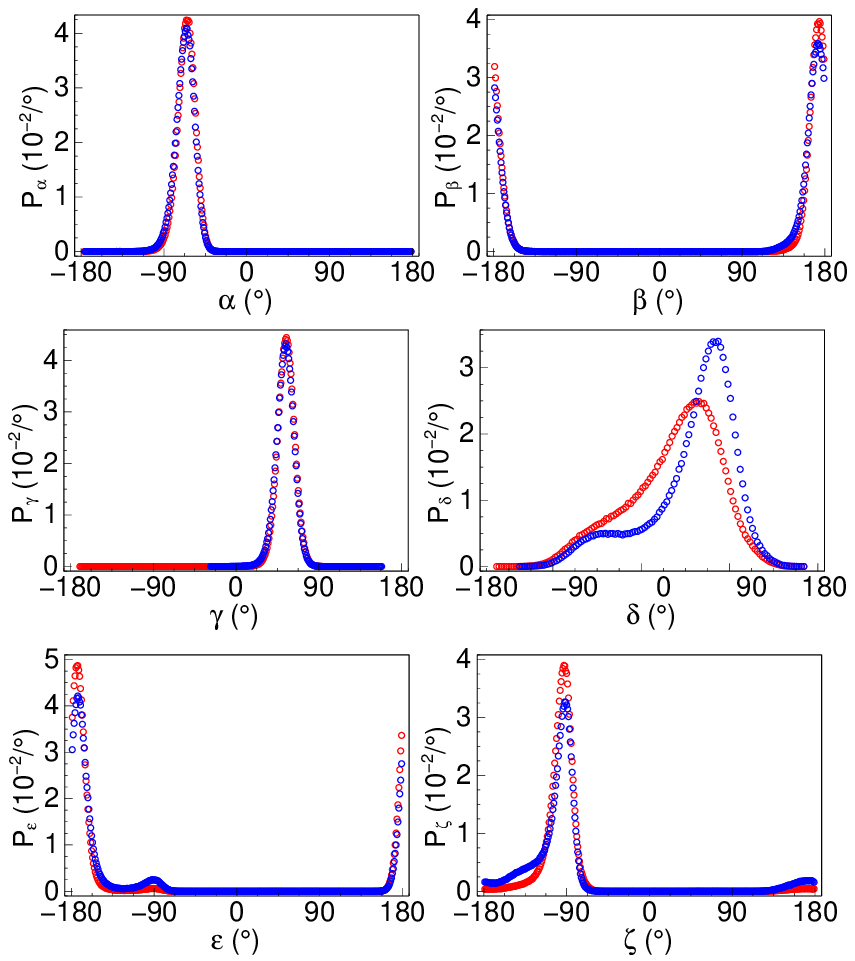}}
\caption{ FIG. \rfg{mdpbd4}S: \footnotesize\label{Fmdpbd4}
Distributions of backbone torsion angles in free MD simulations of
d(ApT)$_7$ (red circles) and d(GpC)$_7$ (blue circles).
}\end{figure}%=====================================================

In all simulations the B-DNA conformations were well conserved
without signs of accumulated deformations. Representative
probability distributions of backbone torsions are shown in
\Rfg{mdpbd4}. They correspond to standard B-DNA dynamics
\cite{Dixit:05a}.

\subsection*{Evaluation of statistical errors}

Evaluation of errors in MD simulations is based upon the following
assertions from the probability theory. Consider a random variable $x$
with expectation ${\bf M}x=\xi$ and variance ${\bf D}x=\sigma^2$. We
can take $n$ samples of $x$ and compute
$$
\overline{x}=\frac1n(x_1+x_2+...+x_n)=\frac1n\sum^n_{k=1}x_k
$$
and
$$
S^2=\frac1{n-1}\sum^n_{k=1}(x_k-\overline{x})^2
$$
called the sample average and variance, respectively. Both
$\overline{x}$ and $S^2$ are random variables, with ${\bf
M}\overline{x}=\xi$ and ${\bf M}S^2=\sigma^2$, i.e. $\overline{x}$
and $S^2$ provide unbiased estimates of $\xi$ and $\sigma^2$,
respectively. It is also known that
\beq\label{EDx1}
{\bf D}\overline{x}=\frac{\sigma^2}n
\eeq
and, if $x$ is a Gaussian random variable,
\beq\label{EDS}
{\bf D}S^2=2\sigma^4/(n-1).
\eeq
\Req{Dx1} and \req{DS} are used for evaluation of statistical
errors.

Consider evaluation of torsional fluctuations, for instance.  In
this case, the random variable is the twist angle of one helical
turn, $\Phi$, with expectation ${\bf M}\Phi$ and variance ${\bf
D}\Phi$. The torsional persistence length is computed as $l_t=L/{\bf
D}\Phi$. The canonical moments ${\bf M}\Phi$ and ${\bf D}\Phi$ are
thermodynamic limits and they are estimated by using, respectively,
the sample average and variance computed over all $n$ points saved
during an MD trajectory.  However, \Req{Dx1} and \req{DS} cannot be
applied straightforwardly because they are valid only for
statistically independent samples, i.e. the time intervals between
the MD states must be suitably large compared to the torsional
relaxation time.  The data saving interval is commonly much smaller,
therefore, the errors are evaluated by using the method of block
averages \cite{Flyvbjerg:89,Frenkel:96}.  The trajectory is
successively divided in $n'=2,4,...,2^{15}$ stretches (blocks) and
the sample variances $S'^2$ are computed by using $n'$ block
averages instead of individual samples. When the blocks are longer
than the torsional relaxation time the block averages are
independent and $S'^2/n'\approx const=\sigma^2/\tilde n$, where
$\tilde n$ is the effective number of independent samples provided
by the trajectory.  This value should be used in place of $n$ in
\Req{Dx1} and \req{DS}.  In practice, it is convenient to draw the
plots of
$$
\Omega(n')=\frac{nS'^2}{n'\sigma^2},
$$
with respect to $\log_2n'$. When statistical independence is reached,
such plots exhibit a plateau with $\Omega\approx n/\tilde n=\tau^c$,
which gives the required estimate of $\tilde n$. Parameter $\tau^c$ is
the effective correlation time measured in trajectory saving steps.

\Rfg{blkerBD} shows how this works for Brownian dynamics (BD).  We
consider eight trajectories from earlier published simulations of a
discrete wormlike chain (WLC) model of 14-mer DNA
\cite{Mzjpc:09,Mzjctc:09}. Each data set involved $2^{15}$ consecutive
configurations saved with a 5 ps interval (about 164 ns in total),
that is exactly as in MD simulations. With $n'$ decreasing, the plots
display emergence of a plateau and the growth of statistical noise.
The black dashed line displays the results for a single eight times
longer trajectory. It is seen that the plateau becomes less noisy, but
its level does not change. This confirms the validity of the above
derivations, i.e. $\tau^c$ is constant and, consequently, \Req{Dx1}
holds. \Rfg{blkerBD} shows that a reasonably accurate estimate of the
plateau value can be obtained with $n'=2^6$ or larger.  On the other
hand, the plateau emerges with the block length $10\tau^c$ or larger,
therefore, the lower estimate of the necessary total duration of
trajectory is $640\tau^c$.

\begin{figure}[ht]
\centerline{\includegraphics[width=6cm]{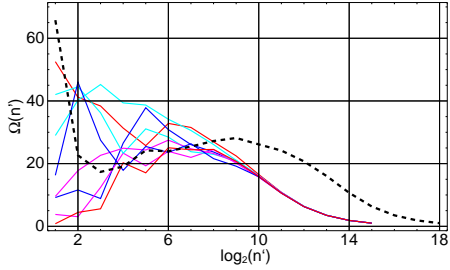}}
\caption{ FIG. \rfg{blkerBD}S: \footnotesize \label{FblkerBD}
Analysis of statistical errors in BD simulations by the method of
block averages \cite{Flyvbjerg:89,Frenkel:96}. The solid color lines
correspond to eight similar 164 ns trajectories. The black dashed line
displays the results for a single eight times longer trajectory.
}\end{figure}%=====================================================

\begin{figure}[ht]
\centerline{\includegraphics[width=6cm]{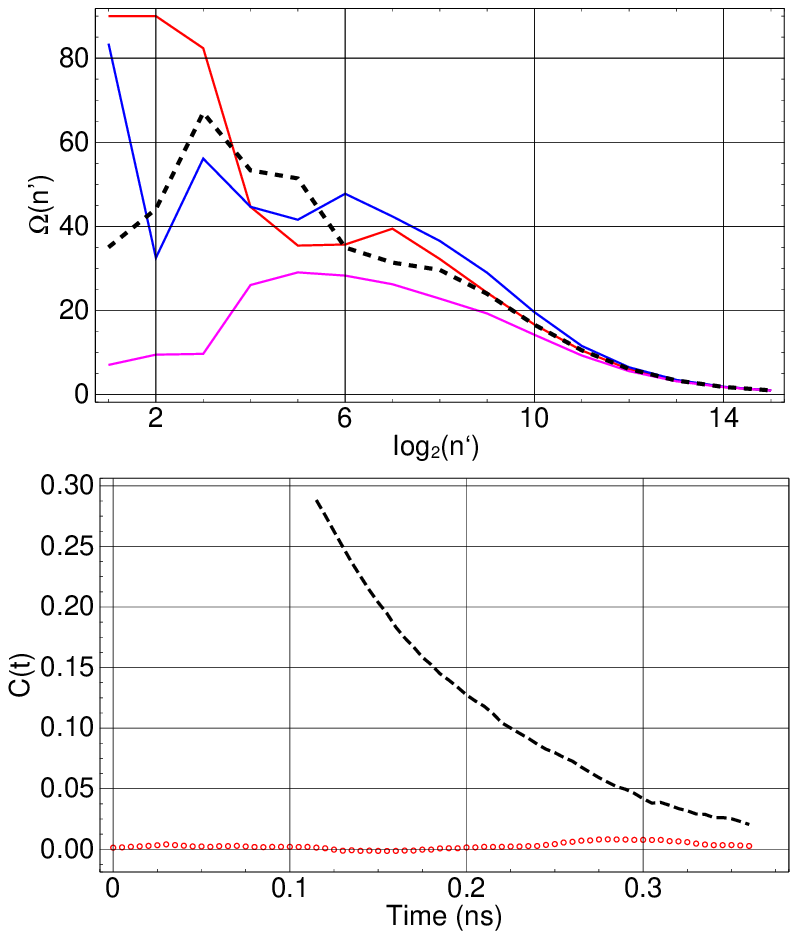}}
\caption{ FIG. \rfg{blkerMD}S: \footnotesize \label{FblkerMD}
Analysis of statistical errors in MD simulations by the method of
block averages \cite{Flyvbjerg:89,Frenkel:96}. In the upper panel, the
solid plots were obtained by concatenating 32 parallel MD
trajectories. The red trace shows the results of free dynamics without
external stress. Analysis of a continuous 164 ns trajectory with the
same total volume of sampling is shown by the black dashed line. The
lower panel shows the time autocorrelation function of the $\Phi$
angle from the continuous trajectory (black dashed trace) and the
corresponding average cross-correlation function computed for 32
parallel trajectories (red circles). The blue and magenta traces in
the upper panel show the results for applied torques of +40 and -40
pN$\cdot$nm, respectively.
}\end{figure}%=====================================================

The foregoing analysis is equally valid when several trajectories are
concatenated and considered as a single longer trajectory. It is only
necessary that the sub-trajectories are statistically independent and
suitably longer than $\tau^c$. A representative example is shown in
the upper panel of \Rfg{blkerMD}. We consider the torsional
fluctuations of CG$_6$ fragments. The dashed black plot was
obtained by processing a continuous 164 ns trajectory form our earlier
report \cite{Mzprl:10}. The solid red line shows the analogous
results for the protocol used here (32 concatenated sub-trajectories).
In both cases no external stress was applied. The plots show good
convergence except for small $n'$ where the statistical noise is high.
The statistical dependence between the subtrajectories is analyzed
by using time cross-correlation functions computed as
\beq\label{ECij}
C^{ij}(t)=\frac{\tilde C^{ij}(t)}{\tilde C^{ij}(0)};\
 \tilde C^{ij}(t)=\langle\Phi^i(\tau+t)\Phi^j(\tau)\rangle_\tau
            - \langle\Phi^i\rangle\langle\Phi^j\rangle
\eeq
where the superscripts $i$ and $j$ refer to individual
sub-trajectories. The lower panel compares the cross-correlation
function averaged  over all $i\ne j$ with the time autocorrelation
function from the long trajectory. The latter is obtained by setting
$i=j$ in \Req{Cij}. This figure confirms that the subtrajectories are
statistically independent from the very beginning. The
$\tau^c\approx40$ estimated from the upper panel gives $\tilde n$=820
and the relative error of 4.5\% in the measured $l_t$ values, which is
sufficient for our purposes. This relaxation time is approximately two
times that estimated from the decay of autocorrelations by using the
condition $C(\tau)=1/e$. A similar consistency is observed for BD
simulations (see \Rfg{blkerBD} and Ref. \onlinecite{Mzjpc:09}).

The upper panel of \Rfg{blkerMD} displays two additional traces
obtained in simulations with applied external torques of $\pm40$
pN$\cdot$nm. The torsional elasticity of the CG$_6$ fragment
changes with the applied torque so that the double helix becomes
softer with untwisting. This gives a noticeable increase in the
torsional relaxation time, and, accordingly, the three traces
corresponding to positive, zero, and negative torques diverge. As a
result, for this DNA fragment, the relative error in the measured
$l_t$ values increases with untwisting.

\section*{Results}

\begin{figure}[ht]
\centerline{\includegraphics[width=8cm]{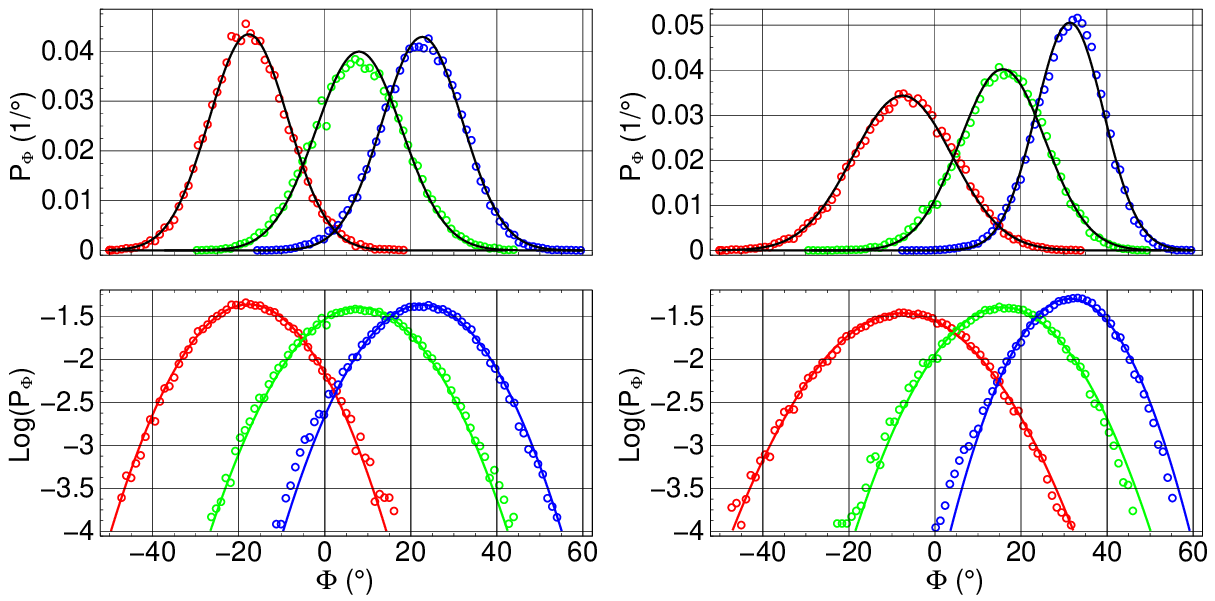}}
\caption{ FIG. \rfg{mdpbd0}S: \footnotesize \label{Fmdpbd0}
The normalized probability density $P_\Phi$ under different
twisting. Data for TA$_6$ and CG$_6$ are exhibited in the left and
right columns, respectively. For each fragment, the results are
displayed for three representative torque values. For TA$_6$, the
plots correspond to $\tau$= -40, +40, and +80 pN$\cdot$nm, from left
to right, respectively. For CG$_6$, the torque values were $\tau$=
-40, 0, and +40 pN$\cdot$nm. The analytical distributions \Req{Pphi}
for to the measured values of $l_t$ and $\Phi_\tau$ are shown by
solid lines. The lower panels display the same data in
semi-logarithmic coordinates.
}\end{figure}%=====================================================

In a harmonic approximation, the canonical distribution of twisting
fluctuations reads
\beq\label{EPphi}
P_{\Phi}\sim\exp\left[-\frac{l_t}{2L}\left(\Phi-\Phi_\tau\right)^2\right].
\eeq
where $\Phi_\tau$ is the shifted equilibrium twist angle under
torque $\tau$. The persistence length $l_t$ defines the width of the
distribution and it must be constant if the harmonic approximation
is valid. The computed patterns of fluctuations appeared
qualitatively similar for all three measured twist angles. In
\Rfg{mdpbd0}, a few representative distributions are exhibited for
fluctuations of angle $\Phi'$.  All of the distributions are close
to the analytical Gaussians defined by \Req{Pphi}, however, their
widths vary in agreement with the observed changes in $l_t$.
Systematic deviations from the Gaussian shape are noticeable only
with the largest positive torques.

\begin{figure}[ht]
\centerline{\includegraphics[width=6cm]{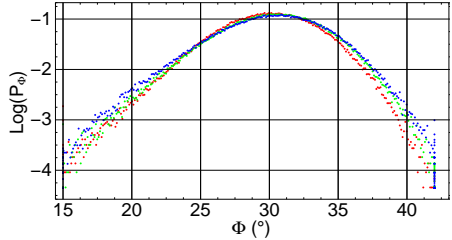}}
\caption{ FIG. \rfg{mdpbd2}S: \footnotesize \label{Fmdpbd2}
The normalized probability densities of twisting fluctuations in the
ApT steps of the TA$_6$ fragment shown in semi-logarithmic
coordinates. The red, green, and blue plots show the data for
external torques of -40, +40, and +80 pN$\cdot$nm, respectively.
}\end{figure}%=====================================================

\begin{figure}[ht]
\centerline{\includegraphics[width=8cm]{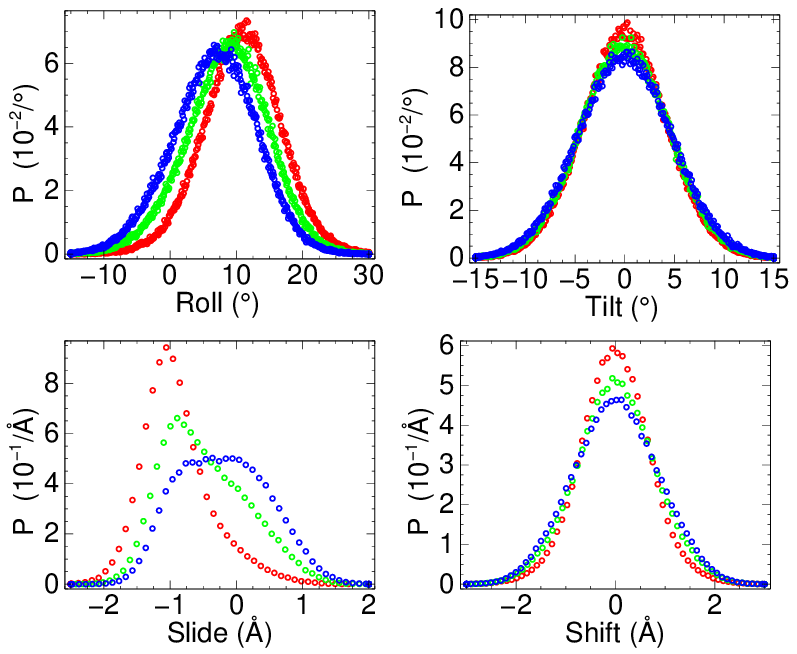}}
\caption{ FIG. \rfg{mdpbd5}S: \footnotesize \label{Fmdpbd5}
The normalized probability densities of fluctuations of base-step
parameters roll, tilt, shift, and slide in TpA dinucleotides of TA$_6$.
The definition of parameters corresponds to the standard convention.
\cite{Olson:01} Their values were computed by the program 3DNA.
\cite{Lu:03c} The red, green, and blue plots show the data for torque
values $\tau$= -40, +40, and +80 pN$\cdot$nm.
}\end{figure}%=====================================================

\clearpage
\bibliography{mzpaper}

\end{document}